# Modelling the error structure in Urban Building Energy Models with a Gaussian Process-based approach

Chunxiao Wang[1, 2, 3], Bruno Duplessis[3], Eric Peirano[1], Pascal Schetelat[2], Peter Riederer[2]
[1]Efficacity, France
[2]CSTB (Scientific and Technical Centre for Building), France
[3]CEEP (Energy, Environment and Processes), Mines Paris – PSL, France

## Abstract

Urban Building Energy Models (UBEM) support urban-scale energy decisions and have recently been applied to use cases requiring dynamic outputs like grid management. However, their predictive capability remains insufficiently addressed, limiting confidence in UBEM application when validation experiments (VE) are unavailable. This study proposes a Gaussian Process (GP)-based method to model the error structure of UBEM, involving: (1) creating a training dataset mapping VE conditions to validation errors, (2) fitting a GP model, and (3) using cross-validation to assess prediction accuracy and uncertainty while extrapolating to unknown scenarios. Applied to the Blagnac (France) district heating network with the UBEM DIMOSIM, GP models effectively capture the inherent structure of UBEM error and uncertainties. Results reveal relationships between model performance and application conditions (e.g., load variation and weather), and show great potential in estimating within-domain model error and extrapolating beyond the validation domain.

## Key Innovations

- Use GP based approach to quantify the error structure of a UBEM,
- Extrapolate the UBEM predictive capability to unvalidated buildings,
- Discover different interaction patterns between validation experiment (VE) conditions and UBEM performance.

## Practical Implications

This paper allows UBEM developpers to have a more comprehensive view of UBEM performance, and helps practionners to prioritise measurement campaigns and to better design further validation experiments.

## Introduction

Urban Building Energy Models (UBEM) have offered a promising solution by enabling informed decision-making at an urban scale. As UBEM methodologies continue to advance, researchers emphasise the need for systematic analysis of model accuracy and bias to improve reliability (Oraiopoulos & Howard, 2022).

Recent developments in UBEM have expanded their application to new use cases, such as grid management and load-shifting studies (Ang et al., 2020). The effectiveness of these tools in such applications depends heavily on their capability to predict accurate dynamic outputs, underscoring the importance of model validation under transient conditions. Imprecise UBEM forecasts can lead to resource misallocation, grid instability, and inefficiencies in load-shifting strategies, thereby undermining both operational reliability and cost-effectiveness.

However, one of the major knowledge gaps in the field of UBEM is the limited focus on their predictive capability. The AIAA Guide (AIAA, 1998) defined *prediction* as "foretelling the response of a system under conditions for which the model has not been validated." Unfortunately, current BEM and UBEM validation studies mostly focus on evaluating model performance under limited validation conditions (Ohlsson & Olofsson, 2021), without attempting to answer to the ultimate question of model validation: how well do models predict in their intended use cases rather than under specific validation conditions? In Figure 1, we can visualise such knowledge gap: a huge discrepancy between the validation domain, which is defined by existing validation experiments (VE), and the application domain, where a model will be used.

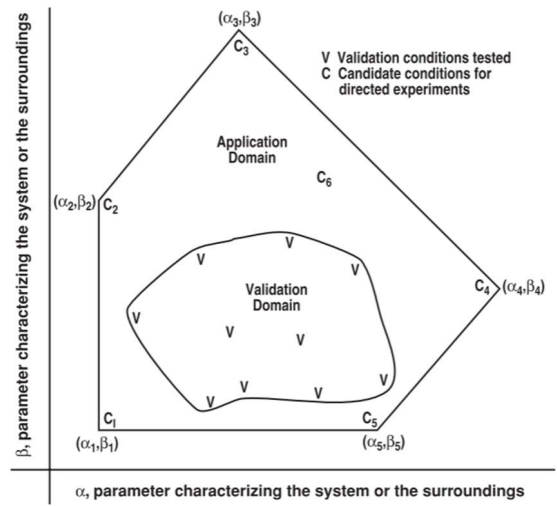

*Figure 1: Illustration of the validation domain and application domain of a model adopted from (Oberkampf & Roy, 2013a).*

As for dynamic predictions of UBEM, one of the main obstacles to narrowing the discrepancy between the validation domain and the application domain is data scarcity: we lack hourly measurements and a variety of building samples. While the first is the basis of any model validation, the second is particularly important to generalise the model's validity from a limited validation domain. Despite the importance of hourly (or sub-hourly) energy consumption data, these time-series are highly sensitive from a privacy standpoint, as they allow to reveal occupant behaviour patterns. Therefore, using them in UBEM validation may run counter to model validation requirements: the heavy use of these data requires significant efforts to address privacy concerns, such as anonymisation (Schaffer et al., 2022), whereas BEM validation experiments require highly accurate model inputs, including geolocated building information and even detailed occupancy profiles. Because of limited access to representative, high-resolution input and measured data, it is difficult to validate UBEMs across the wide range of conditions they might encounter in practice.

Without validation experiments spread across the application domain, the only way to quantify the predictive capability of a model is to extrapolate the structure of model errors to unknown conditions where the model will be used (Oberkampf & Roy, 2013b).

To the best of the authors' knowledge, there is a lack of such attempt in the field of BEM or UBEM. This study is performed to fill this knowledge gap. In this paper, we consider this extrapolation task as a regression problem, seeking to map application conditions to model errors. A Gaussian Process-based approach is chosen for its strong capability to handle small, high-dimensional datasets, naturally quantify predictive uncertainty, and capture the nonlinear relationships inherent in UBEM error structure.

This paper is organised as follows: Section 2 describes the research methodology and explains main methodological choices. Section 3 describes how this methodology is implemented for a case study in Blagnac (France), and how the error structure of the studied UBEM is extrapolated using Gaussian Process (GP) models. Section 4 evaluates the performance of different models, presents the modeled error structure and key findings within this structure. Conclusions are drawn in Section 5.

## General Methodology

As discussed in Section 1, we develop a GP model predicting model error from VE conditions. The general methodology for developing, evaluating and applying this model is structured in 3 steps and shown in Figure 2.

### Generation of VE dataset

The primary goal of this step is to create a set of validation experiment (VE) samples that include both detailed VE conditions and corresponding model errors. Because we aim to extrapolate the model's error structure to unknown scenarios, the chosen VE conditions must be sufficiently representative and contain low uncertainty to ensure that any observed model errors can be explained by their inherent relationship with those conditions.

According to the principles of model validation and verification (V&V), discrepancy due to inaccurate inputs should be minimised. To achieve this, one should use either detailed, accurate inputs or perform model calibration. Ideally, detailed input data can be obtained directly to generate VE samples. However, if the model is instead calibrated to reduce input uncertainty, we suggest to exclude calibration parameters from the error structure model, as calibration of numerous free parameters can introduce additional uncertainty in the quantification of predictive capability (Oberkampf & Roy, 2013a).

Another key requirement is that the recorded validation errors reflect only genuine model errors. According to best practices in BEM validation methodology (Jensen, 1995), all data should allow the identification of discrepancy sources. Consequently, data cleaning is critical—not only to remove measurement errors but also to eliminate specific periods or events, where model validity is not the primary cause of inconsistency between measured and simulated data.

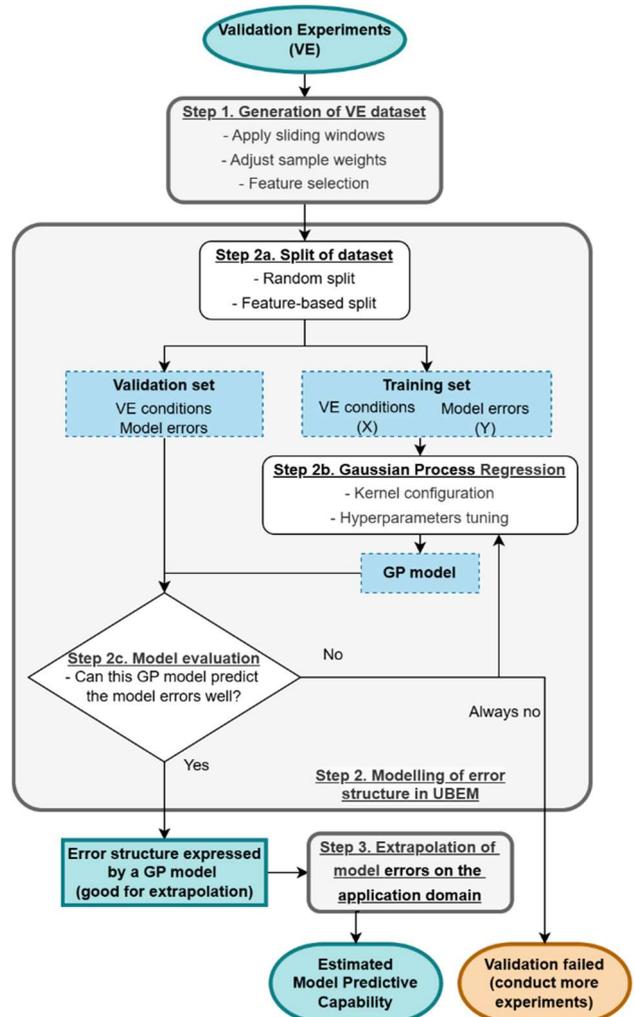

*Figure 2: Research outline.*

### Gaussian Process Regression

In this step, we apply Gaussian Process Regression (GPR) to the VE dataset to investigate the error structure of a UBEM. A Gaussian process is a stochastic process in which every finite collection of variables follows a multivariate normal distribution.

An underlying assumption of this paper is that the distributions of UBEM errors, or their transformations, follow normal distributions under the same VE conditions. This assumption justifies the use of the GPR method in this study.

**Model evaluation**

The key question to be answered here is whether our GP model can predict well UBEM validation errors. The answer made leads to different actions: 1) apply the accurate, robust error structure model to quantify the predictive capablity of UBEM on the application domain, 2) refine GP model, or 3) conduct more validation experiments to obtain more data.

As we are primarily interested in the extrapolation power of the GP model, out-of-sample validation strategies, such as cross-validation, should be prioritised. Thus, the developed model can be evaluate through two tests: interpolation and extrapolation. In the interpolation task, one can apply a random split to VE dataset, to create a training set and validation set. Although GP models never see validation data during training, we still consider this as an interpolation because, if both sets are sufficiently representative samples of the entire VE population, the validation set will follow the same distribution as the training set. Regarding the extrapolation evaluation, we recommend splitting VE dataset based on one or several features to ensure that the extrapolation conditions in the validation set are significantly different from those in the training set on at least one dimension.

For both interpolation and extrapolation tasks, we suggest using at least a probabilistic metric to evaluate the model's performance, such as Negative Log Predictive Density (NLPD), which is negative log likelihood of the test data given the predictive distribution, defined as:

$$NLPD = -\frac{1}{N}\sum_{i=1}^{N} \log p(y_i = t_i | x_i) \qquad (1)$$

where $N$ is the total number of test samples and $p(y_i = t_i | x_i)$ is the model's predicted probability that the outcome $y_i$ will take the observed value $t_i$ given the $i$-th input vector $x_i$. NLPD evaluates the entire predictive distribution, not just a single point prediction, and penalises a model if it is over- or under-confident in its predictions. This makes NLPD the most standard probabilistic metric for evaluating GP models.

**Extrapolation on the application domain**

After validating the GP model through extrapolation tests, we can directly use it to estimate the UBEM's predictive capability on new use conditions. A GP model provides not only error and uncertainty structure within the trained validation domain, but also an estimation of this structure over the application domain where no data are available.

# Case Study

In this paper, VE are conducted for DIMOSIM (Garreau et al., 2021), using measurements collected from a district heating network in Blagnac, France. In this study, we focus on the model capability to predict weekly heating load profiles in apartment buildings. The targeted error metric is then chosen as CV(RMSE), to better represent the model's capability in dynamic predictions.

**Data acquisition**

Energy consumption data in 2021 were collected from buildings connected via thermal substations to the fourth-generation district heating network. The network features 4 kilometres of pipes, 36 substations and a nominal thermal power of 14 MW. One building or small group of buildings is connected to a substation, so 15 substations of apartment buildings are selected as both measurement and validation units. Hourly measurements include average heating power (MW), flow rate (m³/h), and supply and return temperatures (°C), supplemented by monthly energy bills for each substation.

Meteorological data for Blagnac are primarily obtained from the Toulouse-Blagnac airport. Given the station's proximity (only one kilometre away), we assume that the weather data accurately represent local conditions and do not introduce significant bias into our validation.

**Data cleaning**

To ensure that VE outcomes exclusively reflect inherent model errors, we rigorously filter out data issues that are extraneous to model performance. First, low-quality or erroneous measurements are removed using traditional cleaning techniques such as 3-sigma statistical outlier detection, internal consistency checks, and spectral analysis to mitigate noise and quantisation effects. Second, we eliminate data that could introduce additional discrepancies unrelated to model performance. We detect periods of substation malfunction using the method proposed by Gadd and Werner (Gadd & Werner, 2014), and further identify atypical operational patterns by correlating daily energy consumption with heating degree days. Additionally, an anomaly detection is performed using an LSTM-autoencoder, targeting more subtle collective or contextual anomalies that standard methods might miss (Fan et al., 2018). This composite approach ensures that our subsequent validation efforts are both accurate and meaningful.

**Model calibration**

DIMOSIM simulations of substation-level heating loads are calibrated via a brute-force approach. We generates 1,000 random parameter combinations within plausible ranges of all parameters to capture sufficient variability. The calibration was performed in two stages: initially, monthly bills from non-heating periods are used to calibrate domestic hot water usage, and then refining the remaining parameters (e.g., envelope characteristics, temperature setpoint profiles, etc.) by selecting the combination that yields the smallest validation error.

**Multi-period validation**

To increase the effective size of the VE dataset generated from limited measurements, we adopt a rolling-origin approach on two dimensions: calibration period and validation period (Wang et al., 2024). A weekly window is sliced each time, with non-overlapping windows (weekly step) used for calibration periods and overlapping windows (daily step) for validation periods. In each calibration window, the model yielding the smallest error is selected and then applied to all validation periods which are in the same heating season as the calibration period.

To handle missingness issue in time series data, we decide to remove periods with more than one day of missing data for both calibration and validation. As a result, some dates within the measurement period appear more frequently in the VE dataset; therefore, each VE sample's weight is computed as the sum of the reciprocal frequencies of its constituent dates, then normalised so that the total weight across all samples matches the sample count. This approach ensures that the contribution of each sample is adjusted according to the uniqueness of its constituent dates, and effectively handles temporal dependencies in overlapping windows sliced via a rolling-origin approach.

However, since the validation error is calculated with a calibrated model selected by "brute-force", this error may only reflect the optimal potential of model performance, rather than the actual performance with accurate input data. In the absence of accurate input data at the urban scale, we assume that this calibrated error is a reasonable estimate of UBEM performance.

**Generation of VE dataset**

With a multi-period validation process, measurements and simulation outputs in 2021 can be sliced to a set of distinct VEs, while each of them is characterised by a validation error metric and VE features. In total, the resulting VE dataset consists of 32,301 VE.

To ensure the interpretability of models, we mainly select accurate raw features rather than using dimensionality reduction methods. These features have negligible uncertainty and are potentially related to model errors. Selected features can be grouped in three categories: 1) energy use features, 2) boundary condition features, and 3) calibration and validation (C&V) condition features.

Heating load profiles and weather data are transformed into the first two types of features. For these time series data, the minimum, average, median and maximum of each measure over each VE window are extracted as features. We also create some artificial features based on our field knowledge and literature suggestions: e.g., heating degree days and relative power variation metric ($G_a$) (Gadd & Werner, 2013). The latter measures how the heat load pattern in a building is offset from a flat average load profile. In this paper, $G_a$ is adopted to weekly validation periods as:

$$G_{a,weekly} = \frac{\frac{1}{2}\sum_{i=1,j=1}^{168,7}|P_{h,i}-P_{d,j}|}{P_w \cdot 168} \times 100[\%] \quad (2)$$

where $P_h$ is the hourly average heat load (W), $P_d$ is the daily average heat load (W), $P_w$ is the weekly average heat load (W). The third group of features describes how the model was calibrated and validated: e.g. gaps between the calibration period and the validation period in time and heating degree days.

In the current case study, calibrated parameters (e.g., building envelope parameters, operational patterns, etc.) are excluded from the VE dataset as our calibration process includes a great number of free parameters and therefore calibrated values of these parameters are not as trustworthy as other VE features.

**Correlation analysis in VE samples**

In order to fulfil the hypothesis of multivariate normal distribution in a GP, features and validation errors with skewed distributions are all converted to normal distributions using Box-Cox transformation. All attributes in the VE dataset are then scaled to a range between 0 and 1 to simplify future regressions.

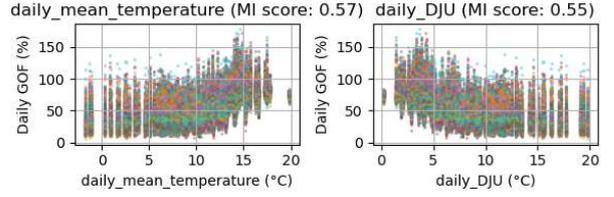

*Figure 3: Mutual information score between daily CV(RMSE) and extracted daily features.*

Before developing GP models, we conduct a correlation analysis among transformed attributes of the VE dataset to establish some prior knowledge about the error structure of the model. We find that the model's error can be monotonically correlated to some VE features with non-linear relationships, as illustrated in Figure 3. Given this correlation nature, we used distance correlation ($dcor$) to measure the association between errors and VE features. For each group of features, we select three most correlated features, excluding highly pairwise-correlated features ($dcor > 0.8$). The list of features is given in Table 1 and detail definitions of these features are in the nomenclature section at the end of this paper. Partial results of the correlation analysis are shown in Figure 4.

*Table 1: Three groups of studied features and their correlation coefficients with transformed CV(RMSE)*

| Feature group | Feature name | dcor |
|---|---|---|
| Energy use | power_variation_boxocx | 0.76 |
| | median_power_per_m2_boxcox | 0.74 |
| | thermoception | 0.5 |
| Boundary | max_temperature | 0.74 |
| | mean_ghi | 0.71 |
| | temperature_variation | 0.56 |
| C&V | max_temp_gap | 0.49 |
| | ghi_gap | 0.48 |
| | temp_var_gap | 0.35 |

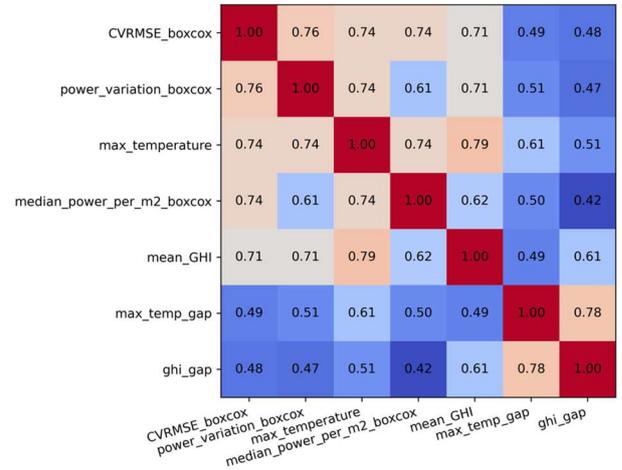

*Figure 4: Distance correlation coefficient matrix.*

**Development of GP models**

Gaussian process regression (GPR) performance can be influenced by several factors, such as kernel choice, hyperparameter tuning, noise characteristics, etc. A number of GP models are developed in this paper using scikit-learn package (Pedregosa et al., 2011), varying in two aspects:
- Size of the training dataset,
- Number of used features.

The default kernel is chosen for GP models: a radial basis function (RBF) kernel representing the stationary assumption, multiplied by a constant kernel (to scale the magnitude of the RBF kernel), and then added to the white kernel that accounts for noise.

In terms of training dataset sizes, we notice that the current VE dataset with over-lapping samples is too large for common GPR task. Therefore, we decide to sample randomly a subset from the original dataset for both the model training and validation task. To evaluate the impact of this parameter, we train models with a sample size ranging from 250 to 3,000, while using three fixed features – the most correlated ones in the 3 groups of VE features.

Regarding the number of features, we fix the sample size to 1500 and use 1 to 9 features to train the GP models. Each model is added with a new feature following two rules: 1) prioritise features in a VE feature group whose number of features is the smallest, 2) select the feature with the highest distance correlation coefficient value.

In our GP development framework, an alpha array is defined to regularise the noise level within the model. Each VE sample is mapped to an alpha element, which is inversely proportional to the logarithmic transformation of the corresponding weight. Rather than inverting the weights directly, such transformation can attenuate the influence of larger weights on the resulting alpha values. With this approach, observations possessing higher weights contribute a lower alpha value, representing greater confidence in observation samples.

**Model evaluation**

For the interpolation, we randomly select a new sample as the validation set (of the same size as the training set) from the remaining portion of the VE dataset, excluding the training data. For the extrapolation, we split our dataset on the "substation" dimension, since we are interested in extrapolating model performance to unseen buildings. A 15-fold cross validation is then performed on the 15 unique substations (building groups) where we collected measurements.

To evaluate models in interpolation and extrapolation tasks, we use a deterministic metric and an interval-based metric in addtion to NLPD: Means Squared Error (MSE) and Coverage for the 95% confidence interval, to evaluate the accuracy of the predicted means and uncertainties.

**Results and Discussions**

This section presents the research results from two aspects. Firstly, the interpolation and extrapolation results of different models are illustrated and compared. Secondly, the inherent relationships between UBEM error and VE features are carefully examined, providing insights into the revealed structure of UBEM errors.

**Evolution of performance with training set size**

Figure 5 demonstrates the evolutions of MSE and NLPD with training set size, from 250 to 3000 samples. We observe that adding more training samples generally stabilises or slightly improves the interpolation performance, whereas the extrapolation performance can remain less stable or show no visible gain.

Such results suggest that GP models trained with smaller training sets may not be able to well capture the actual uncertainty in VE errors, given the smaller Coverage values as shown in Figure 6. In general, simply feeding more training samples is not a good solution to improve model performance, especially in extrapolation tasks.

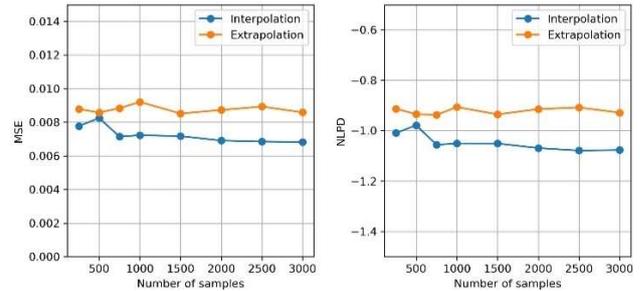

*Figure 5: Evolution of MSE and NLPD on the sample size with fixed features.*

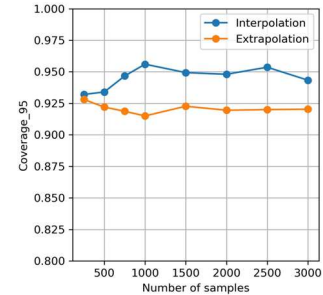

*Figure 6: Evolution of Coverage on the sample size with fixed features.*

**Evolution of performance with number of features**

Figure 7 shows the evolution of model performance with the increasing number of model features. In general, we find that the gap between the interpolation and extrapolation performance of the GP model becomes wider as the number of features increases. In addition, two distinct phases can be observed in these curves: one before and the other after using 4 features.

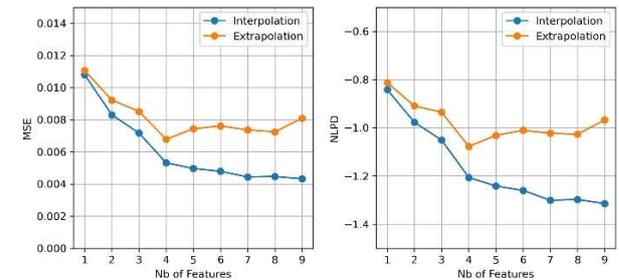

*Figure 7: Evolution of MSE and NLPD on the number of features with fixed sample size.*

The first phase can be characterised by a simultaneous improvement of model performance in both interpolation and extrapolation tasks, as the number of features increases. Despite the model performance gap in these tasks, the trends of evolution remain very similar. However, in the second phase, the performance of the GP model shows very different evolutionary trends in the interpolation and extrapolation tasks: adding more features worsens the extrapolation performance, while the interpolation performance improves much slower than in the first phase.

The reason of the first phase trend is relatively simple: a GP model with more features can capture more relevant variations in UBEM errors, so the increase of feature number is beneficial for GP model training. Regarding the trend in the second phase, it can be viewed as a signal of overfitting because the performance gap between interpolation and extrapolation is increasing with the number of features. From the fourth feature, next selected features are highly correlated with the previous ones. For example, the fifth selected feature, mean Global Horizontal Irradiance (GHI) shares a high distance correlation coefficient of 0.79 with the third selected feature maximum outdoor temperature, while the highest value among previous features is 0.74. This means that the newly added feature brings little new information to the UBEM error structure but results in an overconfident model, indicating a growing underestimation of model uncertainty.

**UBEM predictive capability and VE conditions**

To investigate UBEM's predictive capability under various conditions, we use the model trained with 5 features and 1500 VE samples. Without incurring a high risk of overfitting, the current GP model is able to provide insights into the evolution of UBEM errors across different features, taking their interactions into account.

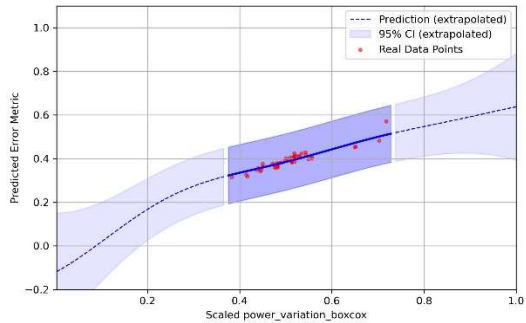

*Figure 8: Evolution of UBEM error with the power variation over the validation domain (purple dark area) and outside the validation domain (purple light area).*

The primary feature significantly influencing model performance is the relative power variation (Ga), as depicted in Figure 8. The error structure indicates that greater power variation is associated with higher UBEM error. The highest power variation cases exhibit large prediction errors and wide confidence intervals, reflecting both the difficulty in modeling highly fluctuating energy use and the GP model's uncertainty where VE samples are sparse. In lower variation cases, we find the lowest level of UBEM error. However, it is challenging to determine whether this "basin" is a good representation of UBEM predictive capability or whether it simply corresponds to scenarios where the UBEM is easier to match with the measurements during model calibration.

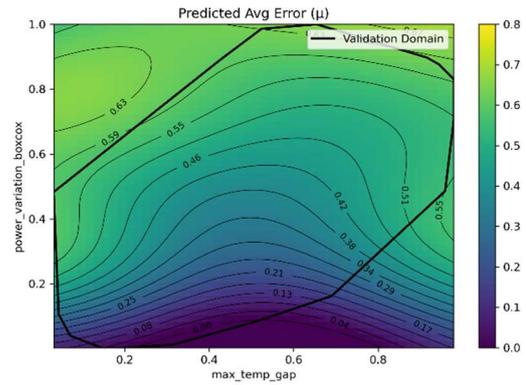

*Figure 9: Heatmap of predicted UBEM error with power variation and gaps of maximum temperature.*

Regarding interactions between different VE features, we notice that common patterns can be grouped into two types: additive and synergistic.

The first pattern, a more common one, suggests the joint effect of two features is simply the sum of their individual contributions. For example, in Figure 9, the relative power variation metric exhibits a monotonical correlation with UBEM error across all maximum outdoor temperature gaps between the calibration and the validation period, whereas extreme values of the latter feature lead to higher UBEM error, indicating poorer predictive capability when calibration and validation conditions differ. Conversely, Figure 10 presents a case in which two features exhibit opposite monotonic correlations with UBEM error. Their simultaneous increase results in a compensatory effect that leaves the UBEM error almost unchanged.

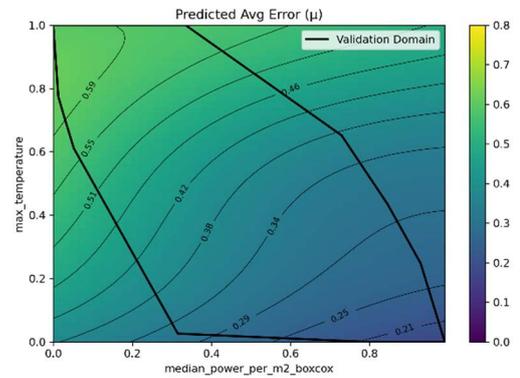

*Figure 10: Heatmap of predicted UBEM error with median power / $m^2$ and maximum outdoor temperature.*

The second type of pattern is rarer and illustrates a synergistic interaction between features. An example between the average GHI and the maximum outdoor temperature is illustrated in Figure 11. At a given level of GHI, there exists a "basin" of minimal UBEM error at an intermediate outdoor temperature. Deviating from this optimal temperature in both directions leads to an increase in prediction error. Although the individual correlations of GHI and outdoor temperature with UBEM error are both monotonic, the effect of outdoor temperature on UBEM error is conditional on the level of GHI. This indicates that both features must be considered simultaneously to fully capture their combined influence on UBEM error. The resulting non-linear relationship underscores the importance of investigating the ensemble influence of features, as the impact of one variable is clearly modulated by the other.

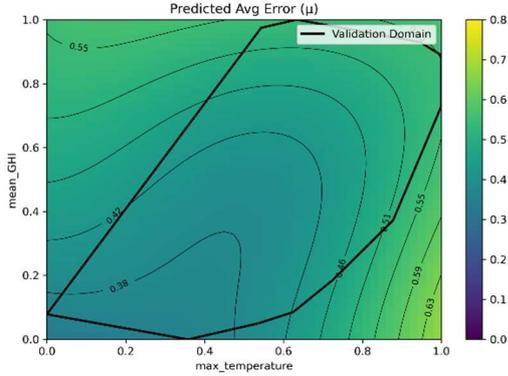

*Figure 11: Heatmap of predicted UBEM error with average GHI and maximum outdoor temperature.*

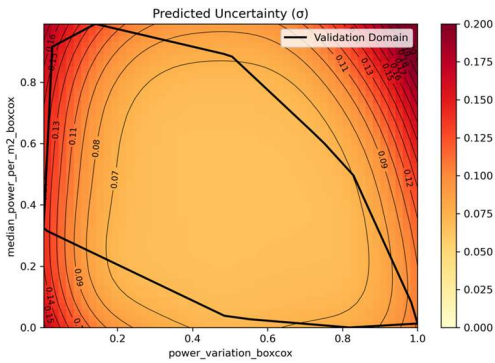

*Figure 12: Heatmap of GP prediction uncertainty (σ) with two features.*

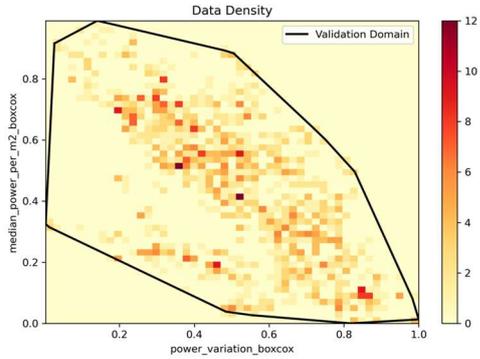

*Figure 13: Heatmap of VE sample density in the validation domain represented in 2D.*

### Uncertainty in predictions of GP models

We observe that the confidence of GP model is associated with the data density over the validation domain: sparser VE data points result in higher prediction uncertainty, as shown in Figure 12 and Figure 13. Meanwhile, at the boundaries of the validation domain (a convex hull defined by extreme VE conditions), the prediction uncertainty is not necessarily as high as in within-domain regions where data are sparse. This observation is quite consistent with how a Gaussian Process handles sparse data: a GP increases predictive variance in sparse regions, reflecting higher uncertainty where data are scarce.

### Extrapolation to unseen buildings

Given previous comparisons between different models, we use the model trained with 4 features and 1500 VE samples to evaluate the performance of GP model while extrapolating to unseen buildings. Table 2 shows the evaluation results for the 15 validation sets (substations).

*Table 2: Model extrapolation performance.*

| Metric | Overall | Min | Max |
|---|---|---|---|
| MSE | 0.007 | 0.004 | 0.011 |
| Coverage | 91% | 78% | 99% |
| NLPD | -1.08 | -1.35 | -0.58 |

Although the overall performance of GP model is statisfying, model's performance varies noticeably across different substations. The GP model clearly handles certain buildings more accurately than others. This implies the heterogeneity in predictive capability of DIMOSIM towards different buildings.

After investigating the modelled error structure, an explanation is found for this performance heterogeneity. In Figure 10, errors are higher when the energy use is lower for a given level of the outdoor temperature, which suggests that buildings with higher energy efficiency are more difficult to predict for DIMOSIM or, ingeneral, to numerical building models.

In fact, when energy-efficient buildings are excluded for model training, the GP model is best trained for energy-inefficient buildings and the extrapolation on efficient ones signifies a domain shift from the previous validation domain to an unseen domain. As this dimension is not considered by the current approach, the prediction is actually made outside the validation domain without specifically informing the GP model. On the other hand, if the model is trained on both inefficient and efficient buildings and validated on a building within this energy efficiency range, this prediction would be more of an interpolation since the model would have already seen similar buildings.

This limitation indicates that the construction of a VE dataset should also integrate building-specific features, such as energy-efficiency indicators. Expandingthe coverage of these features could enhance the extrapolation power of the GP model.

### Conclusion and Perspectives

This study demonstrates that a Gaussian Process (GP) approach can effectively support the assessment of UBEM validity by mapping validation experiment conditions to model errors. Using a VE dataset with carefully selected features, the GP model captures both deterministic trends and inherent uncertainties in the UBEM error structure. Performance evaluations reveal that while the GP model is robust in interpolation tasks—when VE conditions are well represented—it exhibits increased uncertainty and diminished predictive capability when extrapolating to unseen scenarios, especially in regions with sparse data or different building characteristics.

Key findings include the identification of two distinct interaction patterns in the UBEM error structure: additive interactions, where the joint impact of features approximates the sum of their individual contributions, and synergistic interactions, where the influence of one variable is modulated by the level of another. Such insights underscore the importance of simultaneously investigating multiple VE features when assessing UBEM predictive capability.

From a practical perspective, these results offer valuable guidance for future model validation. The GP model's ability to highlight regions of high uncertainty can guide the prioritisation of measurement campaigns and the design of further validation experiments. By targeting under-represented conditions—such as those in more energy-efficient buildings or during extreme weather events—practitioners can incrementally improve the representativeness of VE datasets, yielding a more reliable quantification of UBEM performance even without dedicated validation experiments.

Looking forward, several perspectives emerge from this work. First, expanding the VE dataset to encompass a broader spectrum of building typologies is essential for enhancing the extrapolation power of GP models. Second, integrating additional building-related features may extend the current error structure to new dimensions that are presently challenging to extrapolate. Finally, exploring hybrid or alternative modelling strategies that combine GP with other statistical approaches may help address the limitations observed in current GP models.

## Acknowledgement

This work was supported by Veolia, by providing meter readings from 34 substations from 2021 to 2023.

## Nomenclature: feature explanation

**Prefix**

| | |
|---|---|
| *max_/mean_/median_* | Maximum / mean / median feature value during the validation period |

**Suffix**

| | |
|---|---|
| *_boxcox* | Transformed using Box-Cox transformation |
| *_gap* | Difference between the current feature of the validation period and the calibration period |

**Features**

| | |
|---|---|
| *power_variation* | Defined in Eq. 2 |
| *median_power_per_m2* | Median of hourly average power per unit of floor area |
| *thermoception* | Ratio between the relative hourly average power variation and the absolute temperature variation |
| *max_temperature* | Maximum outdoor temperature |
| *mean_ghi* | Average Global Horizontal |
| *temperature_variation* | Absolute outdoor temperature variation from the average level |
| *max_temp_gap* | Gap in *max_temperature* |
| *ghi_gap* | Gap in *mean_GHI* |
| *temp_var_gap* | Gap in *temperature_variation* |